\documentclass[twocolumn,showpacs,aps,prd]{revtex4-2}
\usepackage{overpic,graphicx}
\usepackage{dcolumn}
\usepackage{bm}
\usepackage{rotating}
\usepackage{subfigure}
\usepackage{color}
\usepackage[bookmarksnumbered, pdfstartview=FitH,colorlinks,urlcolor=blue, citecolor=blue,linkcolor=blue,] {hyperref}
\usepackage{multirow}
\usepackage{amsmath}
\usepackage{makecell}

\newcommand{\gammaetac}{\psi(3686) \rightarrow \gamma \eta_c}

\newcommand{\psip}{\psi(3686)}

\let\oldequation\equation
\let\oldendequation\endequation

\renewenvironment{equation}
  {\linenomathNonumbers\oldequation}
  {\oldendequation\endlinenomath}

\begin{document}

\title{\bf \boldmath
Observation of the hindered electromagnetic Dalitz decay $\psi(3686) \to e^+ e^- \eta_c$
}

\author{
\begin{small}
\begin{center}
M.~Ablikim$^{1}$, M.~N.~Achasov$^{11,b}$, P.~Adlarson$^{70}$, M.~Albrecht$^{4}$, R.~Aliberti$^{31}$, A.~Amoroso$^{69A,69C}$, M.~R.~An$^{35}$, Q.~An$^{66,53}$, X.~H.~Bai$^{61}$, Y.~Bai$^{52}$
% ...
\\
\vspace{0.2cm}
(BESIII Collaboration)\\
\vspace{0.2cm} {\it
$^{1}$ Institute of High Energy Physics, Beijing 100049, People's Republic of China\\
% ...
$^{4}$ Bochum Ruhr-University, D-44780 Bochum, Germany\\
% ...
$^{11}$ G.I. Budker Institute of Nuclear Physics SB RAS (BINP), Novosibirsk 630090, Russia\\
% ...
$^{31}$ Johannes Gutenberg University of Mainz, Johann-Joachim-Becher-Weg 45, D-55099 Mainz, Germany\\
% ...
$^{35}$ Liaoning Normal University, Dalian 116029, People's Republic of China\\
% ...
$^{52}$ Southeast University, Nanjing 211100, People's Republic of China\\
$^{53}$ State Key Laboratory of Particle Detection and Electronics, Beijing 100049, Hefei 230026, People's Republic of China\\
% ...
$^{61}$ University of Jinan, Jinan 250022, People's Republic of China\\
% ...
$^{66}$ University of Science and Technology of China, Hefei 230026, People's Republic of China\\
% ...
$^{69}$ University of Turin and INFN, (A)University of Turin, I-10125, Turin, Italy; (B)University of Eastern Piedmont, I-15121, Alessandria, Italy; (C)INFN, I-10125, Turin, Italy\\
$^{70}$ Uppsala University, Box 516, SE-75120 Uppsala, Sweden\\
% ...
$^{b}$ Also at the Novosibirsk State University, Novosibirsk, 630090, Russia\\
% ...
}\end{center}

\vspace{0.4cm}
\end{small}
}

\begin{abstract} 

Using a data sample of $(448.1 \pm 2.9)\times10^6 ~\psip$ decays collected at an $e^+ e^-$ center-of-mass energy of $3.686~\rm{GeV}$ by the BESIII detector at BEPCII, 
we report an observation of the hindered electromagnetic Dalitz decay $\psi(3686) \to e^+ e^- \eta_c$ with a significance of $7.9\sigma$. 
The branching fraction is determined to be $\mathcal{B}(\psi(3686) \to e^+ e^- \eta_c) = (3.77 \pm 0.40_{\rm stat.} \pm 0.18_{\rm syst.})\times 10^{-5}$, agreeing well with the prediction of the vector meson dominance model. 
This is the first measurement of the electromagnetic Dalitz transition between the $\psi(3686)$ and the $\eta_c$, 
which provides new insight into the electromagnetic properties of this decay, 
and offers new opportunities to measure the absolute branching fractions of $\eta_c$ decays.

\end{abstract}


\maketitle

\oddsidemargin  -0.2cm
\evensidemargin -0.2cm

\section{INTRODUCTION}
The electromagnetic (EM) Dalitz decays $V \rightarrow \ell^+ \ell^- P$ arise from an internal conversion of the virtual photon ($\gamma^*$) in the corresponding decays $V \rightarrow \gamma^{*} P$. 
These decays can be used to probe the dynamic EM structure of the transition $V\to P$, 
and investigate the fundamental mechanisms for the interactions between photons and hadrons~\cite{Landsberg:1986fd}. 
Here, $V$ and $P$ are vector and pseudoscalar mesons, and $\ell$ denotes leptons $(\ell =e,~\mu)$, respectively. 
Normalized to the width of the corresponding radiative decay $V\rightarrow \gamma P$, the differential decay width of $V \rightarrow  \ell^+ \ell^- P$ is described by the formula: 
\begin{eqnarray}
\label{equ:dGVPll}
	\frac{{\rm{d}} \Gamma\left( V \rightarrow \ell^{+} \ell^{-} P \right)}{{\rm{d}} q^{2} \Gamma(V \rightarrow  \gamma P)}
&=&\left|\frac{f_{V P}\left(q^{2}\right)}{f_{V P}(0)}\right|^{2} \times {\rm{QED}}(q^{2}) 
\end{eqnarray}
where $q^2$ is the squared four-momentum transfer that equals the square of the invariant mass of the lepton pair ($M^2_{\ell^+ \ell^-}$), 
$f_{V\!P}(q^{2})$ is the transition form factor (TFF) that characterizes the EM structure of the region in which $V$ converts into $P$, $F_{VP}(q^{2})\equiv f_{V\!P}(q^{2})/f_{V\!P}(0)$ is the normalized TFF with a normalization of $F_{V\!P}(0)=1$, 
and QED$(q^2)$ represents the quantum electrodynamics (QED) calculations of the differential decay width of $V\to \ell^{+} \ell^{-} P$ assuming $V$ and $P$ to be point-like particles~\cite{Kroll:1955zu}.  

According to Eq.~(\ref{equ:dGVPll}), the $q^2$-dependent $f_{V\!P}(q^{2})$, or $F_{VP}(q^{2})$ can be extracted from the experimental data after the QED factors have been taken into account. 
It can also be calculated theoretically using nonperturbative QCD models~\cite{Achasov:1990at, Klingl:1996by, Faessler:1999de, Terschluesen:2010ik, Ivashyn:2011hb}. 
For example, the vector meson dominance (VMD) model can describe the coupling of $\gamma^{*}$ to $V$ via an intermediate virtual vector meson $\mathcal{V}$. 
This mechanism is appropriate in the time-like $q^2$ region, $(2m_{\ell})^2<q^2<(m_{V}-m_{P})^2$, 
where the resonant behavior of $\gamma^*$ arises near $q^2=m^2_{\mathcal{V}}$ since $m_{\mathcal{V}}$ is approaching or even reaching the mass shell~\cite{Landsberg:1986fd}. 
Under a single pole approximation, the TFF can be parametrized with the formula,
\begin{eqnarray}
\label{equ:formfactor}
F_{V\!P}(q^{2}) = \frac{1}{1 - q^{2}/\Lambda^{2}},
\end{eqnarray}
where $\Lambda$ serves as an effective pole mass, subsuming effects from all possible resonance poles and scattering terms in the time-like kinematic region. 
The VMD assumption has been phenomenologically very successful in describing the experimental TFF behaviors for many analogous EM Dalitz decays, 
such as $\eta^{(\prime)} \to \ell^+ \ell^- \gamma/\omega$~\cite{eta_to_mummugam_spec, eta_to_mumugam_na60_2009, eta_to_mumugam_na60_2016, eta_to_eegam_a2, etap_to_eegam_bes3, bes3_etap_to_eeomega}, 
$\phi \to e^+ e^- \pi^0/\eta$~\cite{kloe_phi_to_pi0ee, kloe_phi_to_etaee}, 
$J/\psi (\psi(3686)) \to e^+ e^- \pi^0/\eta^{(\prime)}$~\cite{bes3_psi_to_eepi0, bes3_psi_to_eeetap},
$\psi(3686)\to e^+ e^- \chi_{cJ}$~\cite{bes3_psip_to_eejpsi}, 
and 
$D^{*0}\to e^+ e^- D^{0}$~\cite{bes3_dsttoeed0}. On the other hand, one case where VMD dramatically fails is in the TFF of the $\omega\to \mu^+ \mu^- \pi^0$ decay~\cite{omega_to_mmpi0_1981, eta_to_mumugam_na60_2009, eta_to_mumugam_na60_2016}. 
Neither the VMD model nor other models~\cite{tershlusen_2010, tershlusen_2012, schneider_2012, shekhter_2003, fuchs_2005} can explain the steep rise of the experimental $\omega$ TFF in a larger $q^2$ region that shows a relative increase close to the kinematic cut-off by a factor of $\sim$10~\cite{eta_to_mumugam_na60_2016}. 
This discrepancy of the $\omega$ TFF between theoretical calculations and experimental measurements motivates our experimental investigation of other $V\to \ell^+ \ell^- P$ decays, 
such as the hindered electromagnetic Dalitz decay $\psi(3686)\to e^+ e^- \eta_c$, 
where `hindered' means the transition matrix element between the $\psi(3686)$ and the $\eta_c$ would vanish in the limit of zero virtual photon energy because of the orthogonality of the wave functions with different principal quantum number~\cite{Sucher:1978wq, quarkonia_2008}. 

Based on the VMD model, the branching fraction (BF) is predicted to be $\mathcal{B}(\psi(3686)\to e^+ e^- \eta_c) = (3.04 \pm 0.45)\times10^{-5}$~\cite{Gu:2019qwo, Zhang:2019xia}, after considering the effects of the polarization of the $\psi(3686)$ produced in $e^+ e^-$ collisions. 
Comparing the theoretical and experimental BFs and $q^2$-dependent TFFs can enrich our understanding of the nature of the $\psi(3686)$ meson, for example by distinguishing between a pure $S$-wave state and an $S$- and $D$-wave mixture~\cite{rosner_sd_mix}. 
In addition, the $\psi(3686)\to e^+ e^- \eta_c$ decay makes it possible for absolute measurements of $\eta_c$ decays.
 
	In this Letter, we present the first experimental study of the EM Dalitz decay $\psip \to e^+ e^- \eta_c$. We analyze $(448.1\pm2.9)\times10^6$ $\psi(3686)$ decays~\cite{Ablikim:2017wyh} taken at a center-of-mass energy of $\sqrt{s}=3.686$ GeV with the BESIII detector. 

\section{DETECTOR AND MONTE CARLO SIMULATIONS}
The BESIII detector is a magnetic
spectrometer~\cite{Ablikim:2009aa} located at the Beijing Electron
Positron Collider (BEPCII). The
cylindrical core of the BESIII detector consists of a helium-based
 multilayer drift chamber (MDC), a plastic scintillator time-of-flight
system (TOF), and a CsI(Tl) electromagnetic calorimeter (EMC),
which are all enclosed in a superconducting solenoidal magnet
providing a 1.0~T magnetic field. 

The simulated data samples are produced using a {\sc geant4}-based~\cite{geant4} Monte Carlo (MC) package that incorporates the geometric description of the BESIII detector and the detector response. 
An inclusive MC sample containing $5.06\times10^8$ inclusive $\psi(3686)$ decays is used to investigate the potential backgrounds. 
The detection efficiency is determined by a signal MC sample of $\psi(3686)\to e^+ e^- \eta_c$ events with inclusive $\eta_c$ decays, and using a TFF parameterized by Eq.(\ref{equ:formfactor}) with $\Lambda=3.773$ GeV/$c^2$~\cite{Gu:2019qwo}. 
The production of the $\psi(3686)$ resonance is simulated by the MC event generator {\sc kkmc}~\cite{ref:kkmc}, where the beam energy spread and initial-state radiation (ISR) in the $e^+ e^-$ annihilation have been taken into account. 
The known decay modes are generated by {\sc evtgen}~\cite{ref:evtgen, ref:evtgen2} utilizing BFs taken from the Particle Data Group~\cite{pdg}, and the remaining unknown decays are modelled with {\sc lundcharm}~\cite{ref:lundcharm, ref:lundcharm2}. 
  
\section{EVENT SELECTION}
To select the signal process, we first save all $e^+ e^-$ combinations in each event, and then search for the $\eta_c$ signal in the recoiling system of each combination.
The BF of $\psip \to e^+ e^- \eta_c$ is determined by
\begin{eqnarray}
  \mathcal{B_{\rm{sig}}} = \frac{N_{\rm{sig}}^{\rm{obs}}} { N_{\psip} \cdot \varepsilon_{\rm{sig}}},
\label{equ:branching_2}
\end{eqnarray}
where $N_{\rm{sig}}^{\rm{obs}}$ is the signal yield obtained through a fit to the recoil mass distribution of $e^+e^-$ ($RM_{e^+ e^-}$), $N_{\psip}$ is the total number of $\psi(3686)$ decays, and $\varepsilon_{\rm{sig}}$ is the detection efficiency.

The decay chain of interest is $\psi(3686)\to e^+ e^- \eta_c$, where the $\eta_c$ decays inclusively. 
Charged tracks detected in the MDC are required to have a polar angle ($\theta$) satisfying $|\cos \theta|<0.93$ with respect to the positron beam, 
and a distance of closest approach to the interaction point within $\pm 10$~cm along the beam direction and $1$~cm in the plane transverse to the beam direction. 
The number of good charged tracks is at least 2 with a net charge of 0. Positron (electron) PID uses the measured information in the MDC, TOF and EMC. 
The combined likelihoods ($\mathcal{L}$) under the positron (electron), pion, and kaon hypotheses are obtained. 
Positron (electron) candidates are required to satisfy $\mathcal{L}(e)>0.001$ and $\mathcal{L}(e)/(\mathcal{L}(e)+\mathcal{L}(\pi)+\mathcal{L}(K))>0.8$. 
Within each event, at least one $e^+ e^-$ pair is selected, and the momenta of $e^{\pm}$ should be less than 0.8 GeV/$c$. 
If there are multiple candidates, all of them are kept. 

To reject events from the $\psi(3686) \to \pi^+ \pi^- J/\psi$ decay, each charged track is treated as a pion and the recoil mass of each pair of oppositely charged tracks is required to be outside a range of $[3.090, 3.104]$ GeV/$c^2$. 
Background events from the process $\pi^0(\eta) \rightarrow \gamma e^+ e^-$ are also possible since the recoil mass of the $e^+ e^-$ pair may lie within the signal range of $\psi(3686) \rightarrow e^+ e^- \eta_c$. Therefore, we combine each $e^+ e^-$ pair with a soft photon to form the invariant mass $M_{\gamma e^+ e^-}$. 
The soft photon candidates are identified using showers in the EMC.  The deposited energy of each shower must be more than 25~MeV in the barrel region ($|\cos \theta|< 0.80$) and more than 50~MeV in the end cap region ($0.86 <|\cos \theta|< 0.92$). To suppress electronic noise and showers unrelated to the event, the difference between the EMC time and the event start time is required to be within [0, 700]~ns.
The resulting $M_{\gamma e^+ e^-}$ for each $\gamma e^+ e^-$ combination is required to be outside the ranges $[0.115, 0.150]$ and $[0.505, 0.570]$ GeV/$c^2$ to veto $\pi^0$ and $\eta$ backgrounds, respectively.

\section{BACKGROUND AND SIGNAL YIELD}
The radiated photon in the $\gammaetac$ decay may convert into an $e^+ e^-$ pair when it interacts with the beam pipe or the MDC inner wall, which may result in a fake signal. 
In order to suppress such background, we veto events in which the distance from the reconstructed vertex point of the $e^+ e^-$ pair to the beam direction ($R_{xy}$) is larger than $2~\rm{cm}$ and the angle between the $e^+$ and $e^-$ ($\theta_{e^+ e^-}$) is larger than $40^{\circ}$. 
Here, $40^{\circ}$ is determined by maximizing the figure-of-merit, $S/\sqrt{S+B}$ with respect to $\theta_{e^+ e^-}$, where $S$ and $B$ denote the expected signal yield and the background yield, respectively. 
MC-simulation studies show that the requirement of $\theta_{e^+ e^-}<40^{\circ}$ rejects about 15\% of the signal, but suppresses about 60\% of the background.

%

\begin{figure}[htbp]
	\includegraphics[width=0.45\textwidth]{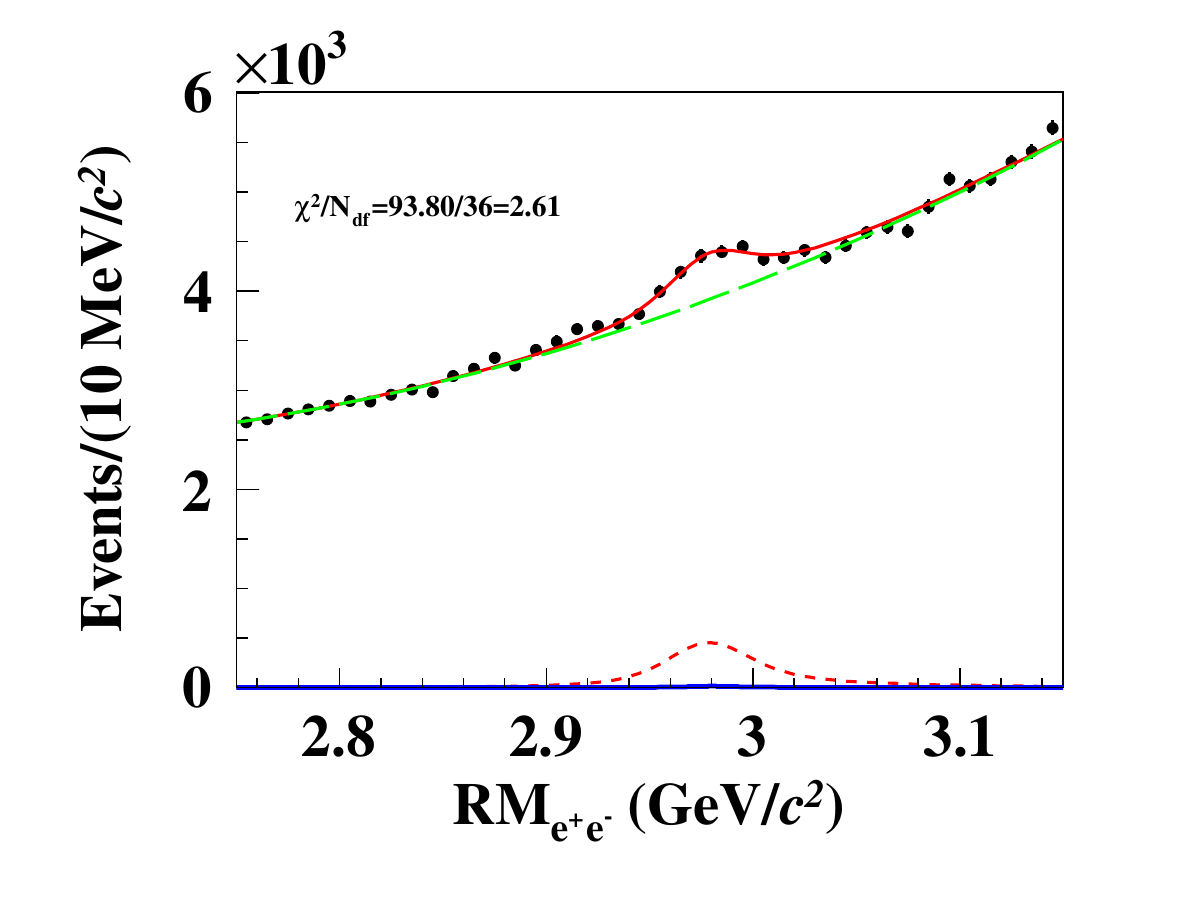}
  \caption{The $RM_{e^+ e^-}$ distribution with the fit results overlaid. The dots with error bars are from data, the solid red line is the best fit, the long dashed green line is the non-peaking background, the solid blue line is the peaking background with a fixed size of 76, and the dashed red line is the signal.}
  \label{fig:plot_fit}
\end{figure}

\begin{figure}[htbp]
  \includegraphics[width=0.45\textwidth]{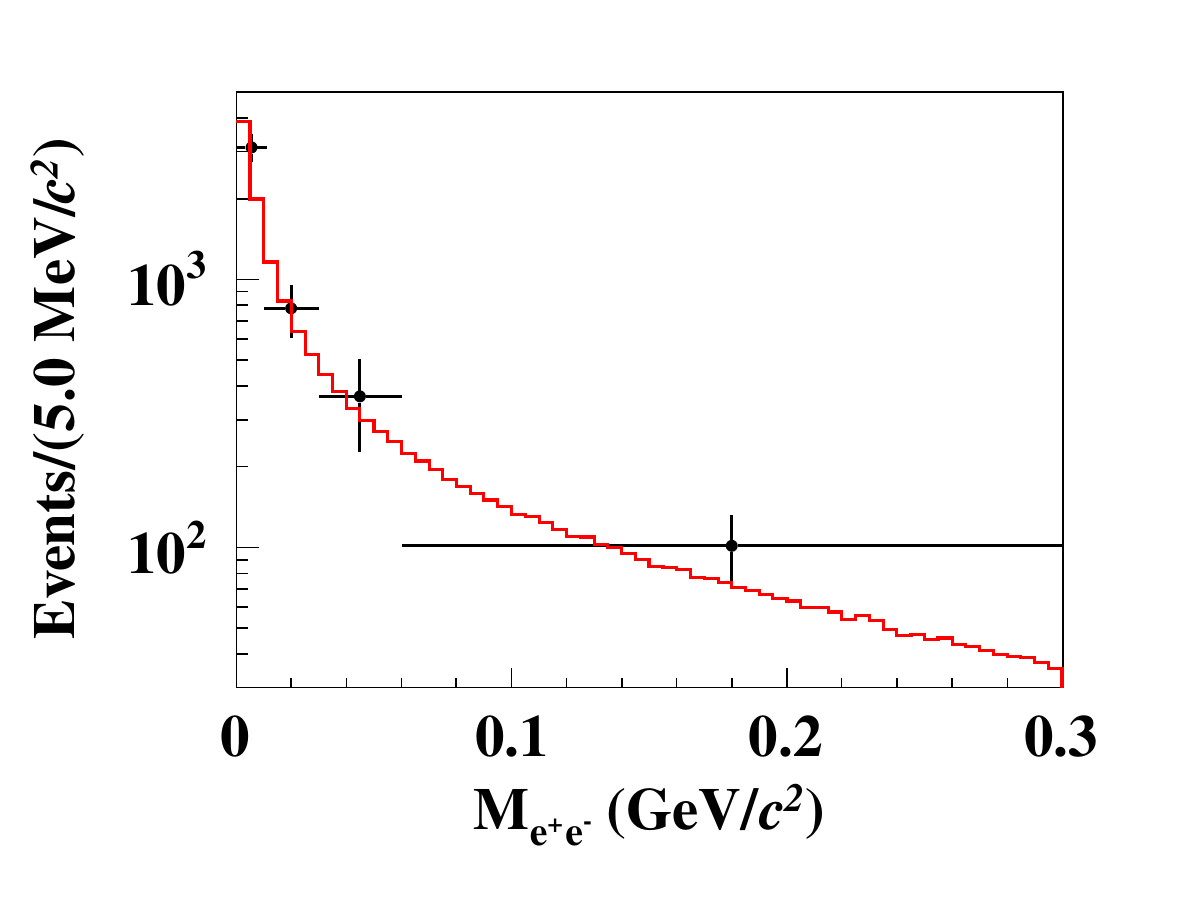}
	\caption{Efficiency-corrected signal yields versus $M_{e^+ e^-}$. The solid red histogram with a bin width of 5 MeV$/c^2$ shown in y-axis is from the signal MC sample, and the four dots with error bars are from data due to the limited statistics, which have been averaged by a bin width of 5~${\rm MeV}/c^2$. Here the four bins are $[2m_e, 0.01)$, $[0.01, 0.03)$, $[0.03, 0.06)$, $[0.06, 0.30]$ GeV$/c^2$. The signal yields in data are obtained from fits to $RM_{e^+ e^-}$ in each bin, and the signal yields in the signal MC sample are normalized by the BF of $\psi(3686) \to e^+ e^- \eta_c$ measured in this letter.}
  \label{fig:plot_mee}
\end{figure}

After applying the above event selection criteria, a clear $\eta_c$ signal peak appears in the $RM_{e^+ e^-}$ distribution, as shown in Fig.~\ref{fig:plot_fit}. 
An unbinned maximum likelihood fit to the $RM_{e^+ e^-}$ distribution is performed to obtain the signal yield. 
In the fit, the signal shape is described by a signal MC shape convolved with a Gaussian function, 
which is used to compensate for the discrepancy in the detection resolution between data and MC simulation. 
The shape of the non-peaking background is modelled with a second order Chebyshev polynomial function, and the size is allowed to float. 
The peaking background contains two contributions. One is from the two-photon process $\gamma^* \gamma^* \to \eta_c$, which is investigated using the $(2916.94 \pm 29.17)$ pb$^{-1}$~\cite{Ablikim:2013ntc, BESIII:2015equ} data sample taken at $\sqrt{s}=3.773$ GeV by BESIII. Here the two virtual photons are radiated from the initial $e^+$ and $e^-$, respectively.
With the same event selection criteria, the number of events for $\gamma^* \gamma^* \to \eta_c$ process ($N^{\gamma \gamma}$) is determined to be $N^{\gamma \gamma} =0^{+67}_{-0}$. 
%
Normalized with the luminosity and the cross section~\cite{Asner:2008nq,twophoton_init}, it is estimated to be $N^{\gamma \gamma} = 0^{+14}_{-0}$ in the $\psi(3686)$ data.
The other contribution is from $\gamma$ conversions within the $\psi(3686)\to \gamma \eta_c$ decay. An MC sample including $1\times10^7$ $\psi(3686)\to \gamma \eta_c$ events has been generated and is found to be consistent with data. The $\gamma$ conversion background is studied using this MC sample. Normalized with $\mathcal{B}(\psi(3686)\to \gamma \eta_c)$~\cite{pdg}, the number of events ($N^{\gamma \eta_c}$) in the $\psi(3686)$ data is computed to be $N^{\gamma \eta_c}=76\pm13$. 
Thus, the total number of events for the peaking background is $76^{+19}_{-13}$, where the uncertainty is the quadratic sum of the uncertainties for the two parts, and will be taken into account in the systematic uncertainty. 
The shape of the peaking background is extracted from the MC sample of $\psi(3686)\to \gamma \eta_c$, and the size is fixed to be 76. The fit result is depicted in Fig.~\ref{fig:plot_fit}. 
The signal yield determined from the fit is $N_{\rm sig}^{\rm obs}=3078\pm329$. The detection efficiency is estimated to be 18.22\% using the signal MC simulation. The efficiency-corrected signal yields versus $M_{e^+ e^-}$ for data and MC sample are shown in Fig.~\ref{fig:plot_mee}, which are consistent with each other.

The statistical significance of the $\eta_c$ signal is computed to be $9.5\sigma$ with the formula of $\mathcal{S}=\sqrt{-2\ln(\mathcal{L}_0/\mathcal{L}_{\rm max})}$, where $\mathcal{L}_{\rm max}$ and $\mathcal{L}_0$ are the likelihood values when $N^{\rm obs}_{\rm sig}$ is left free and fixed at 0 in the fit, respectively.

\section{BRANCHING FRACTION}
According to Eq.~(\ref{equ:branching_2}), the BF of $\psip \to e^+ e^- \eta_c$ is measured to be $\mathcal{B}(\psip \to e^+ e^- \eta_c) = (3.77 \pm 0.40_{\rm stat.} \pm 0.18_{\rm syst.}) \times 10^{-5}$, where the first and second uncertainties are statistical and systematic uncertainties, respectively. The systematic uncertainties are elaborated upon below.

\section{SYSTEMATIC UNCERTAINTIES}
The systematic uncertainty for the BF measurement comes from the $e^{\pm}$ tracking and PID efficiencies, the requirements $R_{xy}<2$ cm and $\theta_{e^+ e^-}<40^{\circ}$, vetos of the $\pi^0$, $\eta$ and $J/\psi$ backgrounds, the shape of the non-peaking background, the size of the peaking background, the signal model, and the total number of $\psi(3686)$ events.

The systematic uncertainty from the $e^{\pm}$ tracking and PID efficiency is investigated by analyzing a mixed control sample of radiative Bhabha events $e^+ e^- \to \gamma e^+ e^-$ at $\sqrt{s}=3.686$ GeV and $\psi(3686) \to (\gamma_{\rm ISR}) e^+ e^-$. 
According to the two-dimensional distributions of the polar angle and momentum, the $e^{\pm}$ tracking and PID efficiencies in the control sample are weighted to match those in the signal decay. The data-MC simulation difference of $e^{\pm}$ tracking efficiency is determined to be 0.5\% per track and that of PID efficiency is 0.7\% per track, both of which are assigned as the relevant systematic uncertainty.

The systematic uncertainty associated with the requirement $R_{xy}<2$ cm for the suppression of the $\gamma$ conversion background is estimated to be 1.0\% by using a highly pure sample of $J/\psi \to \pi^{+} \pi^{-} \pi^{0}(\to \gamma e^{+} e^{-})$~\cite{Ablikim:2014nro}. 
The uncertainty due to the requirement $\theta_{e^+ e^-}<40^{\circ}$ is estimated using a control sample of $\psi(3686)\to \gamma \chi_{cJ},~\chi_{cJ}\to e^+ e^- J/\psi(\to e^+ e^-/ \mu^+\mu^-)$ $(J=1,2)$. The difference in the efficiency for the requirement $\theta_{e^+ e^-}<40^{\circ}$ between data and MC is determined to be 1.3\%, which is assigned to be the corresponding systematic uncertainty. 
Utilizing the same control sample, the systematic uncertainty from vetoing $\pi^0(\eta)\to \gamma e^+ e^-$ background is assigned to be 3.0\% through analyzing the invariant mass of the radiative photon and the $e^+ e^-$ pair directly from $\chi_{cJ}$ decays. 
The uncertainty caused by vetoing $J/\psi$ background is computed to be 0.4\% by studying a control sample of $\psi(3686)\to \gamma \chi_{cJ}, ~\chi_{cJ} \to 2(\pi^+ \pi^-)$ $(J=0,1,2)$. 

The systematic uncertainty from the shape of the non-peaking background is estimated by altering the order of the Chebyshev polynomial function from second to third or fourth, and the largest difference on the measured BF, 2.5\%, is taken as the systematic uncertainty. 
The uncertainty due to the size of the peaking background is examined by varying its size within $\pm1\sigma$, and the largest difference in the measured BF, 0.5\%, is associated as the relevant systematic uncertainty.

In the nominal analysis, the signal MC sample is generated using a TFF modelled with a single pole, where $\Lambda$ is fixed at $3.773$ GeV/$c^2$. 
Since the branching fraction is not sensitive to the pole mass, according to the theoretical calculation, 
alternative $\Lambda$ values of 1.5 and 6.0 GeV/$c^2$ are used in the signal MC sample samples, which is a conservative range for the estimation of the systematic uncertainty. 
The average difference in the measured BF is estimated to be 0.9\%, which is taken as the systematic uncertainty from the signal model. The uncertainty from the total number of $\psi(3686)$ events is 0.6\%~\cite{Ablikim:2017wyh}. 

The total systematic uncertainty on the measured BF of $\psi(3686)\to e^+ e^- \eta_c$ decay is computed to be 4.8\% by adding the above systematic uncertainties in quadrature.

The signal significance is estimated again after considering the effects of the assumed background shape, the size of the peaking background and the signal model. Based on different variations, the lowest significance of the $\eta_c$ signal is calculated to be $7.9\sigma$.


\section{SUMMARY}
In summary, we observe the hindered EM Dalitz decay $\psi(3686)\to e^+ e^- \eta_c$ with a significance of $7.9\sigma$ using $(448.1\pm2.9)\times 10^6$ $\psi(3686)$ decays. 
The measured BF is $\mathcal{B}(\psip \to e^+ e^- \eta_c)= (3.77 \pm 0.40_{\rm stat.} \pm 0.18_{\rm syst.}) \times 10^{-5}$, which is consistent with the theoretical prediction from the VMD model~\cite{Gu:2019qwo, Zhang:2019xia} within one standard deviation. 
The ratio over the corresponding radiative decay~\cite{pdg} is determined to be $(1.11\pm0.21)\times10^{-2}$, which is in accord with the QED predictions.  
In the future, with the $3\times10^{9}$ $\psi(3686)$ decays recently collected by BESIII~\cite{bes3_white_paper}, the TFF for $\psi(3686) \to e^+ e^- \eta_c$ can be extracted in the whole $M_{e^+ e^-}$ kinematic range, and absolute BF measurements of $\eta_c$ decays can be performed. 

\section*{\boldmath ACKNOWLEDGMENTS}

The BESIII collaboration thanks the staff of BEPCII and the IHEP computing center for their strong support. This work is supported in part by National Key R\&D Program of China under Contracts Nos. 2020YFA0406300, 2020YFA0406400; National Natural Science Foundation of China (NSFC) under Contracts Nos. 11635010, 11735014, 11835012, 11875054, 11875122, 11935015, 11935016, 11935018, 11961141012, 12022510, 12025502, 12035009, 12035013, 12105077, 12165022, 12192260, 12192261, 12192262, 12192263, 12192264, 12192265; the Chinese Academy of Sciences (CAS) Large-Scale Scientific Facility Program; Joint Large-Scale Scientific Facility Funds of the NSFC and CAS under Contract No. U1832207; 
the CAS Center for Excellence in Particle Physics (CCEPP); 
100 Talents Program of CAS; The Institute of Nuclear and Particle Physics (INPAC) and Shanghai Key Laboratory for Particle Physics and Cosmology; 
Excellent Youth Foundation of Henan Province No. 212300410010; The youth talent support program of Henan Province No. ZYQR201912178; The Program for Innovative Research Team in University of Henan Province No. 19IRTSTHN018; 
ERC under Contract No. 758462; European Union's Horizon 2020 research and innovation programme under Marie Sklodowska-Curie grant agreement under Contract No. 894790; 
German Research Foundation DFG under Contracts Nos. 443159800, Collaborative Research Center CRC 1044, GRK 2149; Istituto Nazionale di Fisica Nucleare, Italy; Ministry of Development of Turkey under Contract No. DPT2006K-120470; National Science and Technology fund; 
National Science Research and Innovation Fund (NSRF) via the Program Management Unit for Human Resources \& Institutional Development, Research and Innovation under Contract No. B16F640076; 
STFC (United Kingdom); 
Suranaree University of Technology (SUT), Thailand Science Research and Innovation (TSRI), and National Science Research and Innovation Fund (NSRF) under Contract No. 160355; The Royal Society, UK under Contracts Nos. DH140054, DH160214; The Swedish Research Council; U. S. Department of Energy under Contract No. DE-FG02-05ER41374.


\begin{thebibliography}{**}
\bibitem{Landsberg:1986fd}
  L.~G.~Landsberg,
  \href{https://www.sciencedirect.com/science/article/abs/pii/0370157385901292?via\%3Dihub}{Phys.\ Rept.\  {\bf 128}, 301 (1985).}


\bibitem{Kroll:1955zu}
  N.~M.~Kroll and W.~Wada, \href{https://journals.aps.org/pr/abstract/10.1103/PhysRev.98.1355}{Phys.\ Rev.\ {\bf 98}, 1355 (1955).}

\bibitem{Achasov:1990at}
N.~N.~Achasov and A.~A.~Kozhevnikov,
\href{https://www.worldscientific.com/doi/abs/10.1142/S0217751X92002180}{Int. J. Mod. Phys. A \textbf{7}, 4825 (1992)} [Sov.\ J.\ Nucl.\ Phys.\  {\bf 55}, 449 (1992)].

\bibitem{Klingl:1996by}
  F.~Klingl, N.~Kaiser and W.~Weise,
  \href{https://link.springer.com/article/10.1007/s002180050167}{Z.\ Phys.\ A {\bf 356}, 193 (1996).}


\bibitem{Faessler:1999de}
  A.~Faessler, C.~Fuchs and M.~I.~Krivoruchenko,
  \href{https://journals.aps.org/prc/abstract/10.1103/PhysRevC.61.035206}{Phys.\ Rev.\ C {\bf 61}, 035206 (2000).}

\bibitem{Terschluesen:2010ik}
  C.~Terschlusen and S.~Leupold,
  \href{https://www.sciencedirect.com/science/article/pii/S0370269310007902?via\%3Dihub}{Phys.\ Lett.\ B {\bf 691}, 191 (2010).}


\bibitem{Ivashyn:2011hb}
 S.~Ivashyn,
 \href{https://vant.kipt.kharkov.ua/ARTICLE/VANT_2012_1/article_2012_1_179.pdf}{Prob.\ Atomic Sci.\ Technol.\  {\bf 2012N1}, 179 (2012).}


\bibitem{eta_to_mumugam_na60_2009} R. Arnaldi {\it et al.} (NA60 Collaboration), \href{https://www.sciencedirect.com/science/article/pii/S0370269309005929?via\%3Dihub}{Phys. Lett. B {\bf677}, 260 (2009).}
\bibitem{eta_to_mumugam_na60_2016} R. Arnaldi {\it et al.} (NA60 Collaboration), \href{https://www.sciencedirect.com/science/article/pii/S0370269316300867?via\%3Dihub}{Phys. Lett. B {\bf 757}, 437 (2016).}
\bibitem{eta_to_mummugam_spec} R. I. Dzhelyadin {\it et al.}, \href{https://doi.org/10.1016/0370-2693(80)90937-5}{Phys. Lett. B {\bf 94}, 548 (1980)}.
\bibitem{eta_to_eegam_a2} P. Adlarson {\it et al.} (A2 Collaboration at MAMI), \href{https://journals.aps.org/prc/pdf/10.1103/PhysRevC.95.035208}{Phys. Rev. C {\bf 95}, 035208 (2017).}
\bibitem{etap_to_eegam_bes3} M. Ablikim {\it et al.} (BESIII Collaboration), \href{https://journals.aps.org/prd/abstract/10.1103/PhysRevD.92.012001}{Phys. Rev. D {\bf 92}, 012001 (2015).}
\bibitem{bes3_etap_to_eeomega} M. Ablikim {\it et al.} (BESIII Collaboration), \href{https://journals.aps.org/prd/pdf/10.1103/PhysRevD.92.051101}{Phys. Rev. D {\bf 92}, 051101(R) (2015).}


\bibitem{kloe_phi_to_pi0ee} A. Anastasi {\it et al.} (KLOE-2 Collaboration), \href{https://doi.org/10.1016/j.physletb.2016.04.015}{Phys. Lett. B {\bf 757}, 362 (2016).}
\bibitem{kloe_phi_to_etaee} D. Babusci {\it et al.} (KLOE-2 Collaboration), \href{https://doi.org/10.1016/j.physletb.2015.01.011}{Phys. Lett. B {\bf 742}, 1 (2015).}
\bibitem{bes3_psi_to_eepi0} M. Ablikim {\it et al.} (BESIII Collaboration), \href{https://journals.aps.org/prd/pdf/10.1103/PhysRevD.89.092008}{Phys. Rev. D {\bf 89}, 092008 (2014).}
\bibitem{bes3_psi_to_eeetap} M. Ablikim {\it et al.} (BESIII Collaboration), \href{https://journals.aps.org/prd/pdf/10.1103/PhysRevD.99.012006}{Phys. Rev. D {\bf 99}, 012006 (2019);} \href{https://journals.aps.org/prd/pdf/10.1103/PhysRevD.104.099901}{Phys. Rev. D {\bf 104}, 099901(E) (2021)}.
\bibitem{bes3_psip_to_eejpsi} M. Ablikim {\it et al.} (BESIII Collaboration), \href{https://journals.aps.org/prl/pdf/10.1103/PhysRevLett.118.221802}{Phys. Rev. Lett. {\bf 118}, 221802 (2017).}
\bibitem{bes3_dsttoeed0} M. Ablikim {\it et al.} (BESIII Collaboration), \href{https://journals.aps.org/prd/pdf/10.1103/PhysRevD.104.112012}{Phys. Rev. D {\bf 104}, 112012 (2021).}

\bibitem{omega_to_mmpi0_1981} R. I. Dzhelyadin {\it et al.}, \href{https://doi.org/10.1016/0370-2693(81)90879-0}{Phys. Lett. B {\bf 102}, 296 (1981).}
\bibitem{tershlusen_2010} C. Terschl{\rm ${\rm \ddot{u}}$}sen, and S. Leupold, \href{https://doi.org/10.1016/j.physletb.2010.06.033}{Phys. Lett. B {\bf 691}, 191 (2010).}
\bibitem{tershlusen_2012} C. Terschl{\rm ${\rm \ddot{u}}$}sen, S. Leupold, and M. Lutz, \href{https://link.springer.com/article/10.1140/epja/i2012-12190-6}{Eur. Phys. J. A {\bf 48}, 190 (2012).}
\bibitem{schneider_2012} S. P. Schneider, B. Kubis, and F. Niecknig, \href{https://journals.aps.org/prd/abstract/10.1103/PhysRevD.86.054013}{Phys. Rev. D {\bf 86}, 054013 (2012).}
\bibitem{shekhter_2003} K. Shekhter, C. Fuchs, A. Faessler, M. Krivoruchenko, B. Martemyanov, \href{https://journals.aps.org/prc/abstract/10.1103/PhysRevC.68.014904}{Phys. Rev. C {\bf 68}, 014904 (2003).}
\bibitem{fuchs_2005} C. Fuchs, A. Faessler, D. Cozma, B. V. Martemyanov, and M. Krivoruchenko, \href{https://doi.org/10.1016/j.nuclphysa.2005.03.125}{Nucl. Phys. A {\bf 755}, 499 (2005).}

\bibitem{Sucher:1978wq}
J.~Sucher,
\href{https://iopscience.iop.org/article/10.1088/0034-4885/41/11/002}{Rept. Prog. Phys. {\bf 41}, 1781 (1978).}
\bibitem{quarkonia_2008}
E. Eichten, S. Godfrey, H. Mahlke, and J. L. Rosner, 
\href{https://journals.aps.org/rmp/pdf/10.1103/RevModPhys.80.1161}{Rev. Mod. Phys. {\bf 80}, 1161 (2008).}


\bibitem{Gu:2019qwo}
L.~M.~Gu, H.~B.~Li, X.~X.~Ma and M.~Z.~Yang,
\href{https://journals.aps.org/prd/abstract/10.1103/PhysRevD.100.016018}{Phys. Rev. D \textbf{100}, 016018 (2019).}

\bibitem{Zhang:2019xia}
J.~Zhang, J.~He, T.~Zhu, S.~Xu and R.~Wang,
\href{https://www.worldscientific.com/doi/abs/10.1142/S0217751X1950129X}{Int. J. Mod. Phys. A \textbf{34}, 1950129 (2019).}

\bibitem{rosner_sd_mix}
J. L. Rosner, \href{https://journals.aps.org/prd/abstract/10.1103/PhysRevD.64.094002}{Phys. Rev. D {\bf 64}, 094002 (2001).}

\bibitem{Ablikim:2017wyh}
M.~Ablikim \textit{et al.} (BESIII Collaboration),
\href{https://doi.org/10.1088/1674-1137/42/2/023001}{Chin. Phys. C \textbf{42}, 023001 (2018).}

\bibitem{Ablikim:2009aa}
M.~Ablikim \textit{et al.} (BESIII Collaboration),
\href{https://www.sciencedirect.com/science/article/pii/S0168900209023870?via\%3Dihub}{Nucl. Instrum. Meth. A \textbf{614}, 345 (2010).}

\bibitem{geant4}
S.~Agostinelli \textit{et al.} ({\sc GEANT4} Collaboration),
\href{https://www.sciencedirect.com/science/article/pii/S0168900203013688?via\%3Dihub}{Nucl. Instrum. Meth. A \textbf{506}, 250 (2003).}

\bibitem{ref:kkmc}
S.~Jadach, B.~F.~L.~Ward and Z.~Was,
\href{https://www.sciencedirect.com/science/article/pii/S0920563200008318?via\%3Dihub}{Nucl. Phys. B Proc. Suppl. \textbf{89}, 106 (2000)}

\bibitem{ref:evtgen}
D.~J.~Lange,
\href{https://www.sciencedirect.com/science/article/pii/S0168900201000894}{Nucl. Instrum. Meth. A \textbf{462}, 152 (2001).}


\bibitem{ref:evtgen2}
R.~G.~Ping,
\href{http://hepnp.ihep.ac.cn/article/doi/10.1088/1674-1137/32/8/001}{Chin. Phys. C \textbf{32}, 599 (2008).}

\bibitem{pdg}
R.~L.~Workman \textit{et al.} (Particle Data Group),
\href{https://academic.oup.com/ptep/article/2022/8/083C01/6651666?login=false}{Prog. Theor. Exp. Phys. {\bf 2022}, 083C01 (2022).}


\bibitem{ref:lundcharm}
J.~C.~Chen, G.~S.~Huang, X.~R.~Qi, D.~H.~Zhang and Y.~S.~Zhu,
\href{https://journals.aps.org/prd/abstract/10.1103/PhysRevD.62.034003}{Phys. Rev. D \textbf{62}, 034003 (2000).}

\bibitem{ref:lundcharm2}
R.~L.~Yang, R.~G.~Ping and H.~Chen,
\href{https://iopscience.iop.org/article/10.1088/0256-307X/31/6/061301}{Chin. Phys. Lett. \textbf{31}, 061301 (2014).}

\bibitem{Ablikim:2013ntc}
M.~Ablikim \textit{et al.} (BESIII Collaboration),
\href{https://iopscience.iop.org/article/10.1088/1674-1137/37/12/123001}{Chin. Phys. C \textbf{37}, 123001 (2013).}

\bibitem{BESIII:2015equ}
M.~Ablikim \textit{et al.} (BESIII Collaboration),
\href{https://www.sciencedirect.com/science/article/pii/S0370269315008990?via\%3Dihub}{Phys. Lett. B \textbf{753}, 629-638 (2016).}
[erratum: \href{https://www.sciencedirect.com/science/article/pii/S0370269320307851?via\%3Dihub}{Phys. Lett. B \textbf{812}, 135982 (2021)}]

\bibitem{Asner:2008nq}
D. M. Asner \textit{et al.}, \href{https://arxiv.org/abs/0809.1869}{Int. J. Mod. Phys. A \textbf{24}, S1 (2009).}

\bibitem{twophoton_init} V. M. Budnev, I. F. Ginzburg, G. V. Meledin, V. G. Serbo, \href{https://www.sciencedirect.com/science/article/abs/pii/0370157375900095?via\%3Dihub}{Phys. Rept. {\bf 15}, 181 (1975).}

\bibitem{Ablikim:2014nro}
M.~Ablikim {\it et al.} (BESIII Collaboration),
  \href{http://dx.doi.org/10.1103/PhysRevD.89.092008}{Phys.\ Rev.\ D {\bf 89}, 092008 (2014).}

\bibitem{bes3_white_paper} M. Ablikim {\it et al.}, \href{https://iopscience.iop.org/article/10.1088/1674-1137/44/4/040001}{Chin. Phys. C {\bf 44}, 040001 (2020).}

\end{thebibliography}
\end{document}